\documentclass[aps,prd,amsmath,amssymb,showpacs,floats,floatfix,nofootinbib,reprint]{revtex4-1}
\usepackage[colorlinks=true, linkcolor=red, citecolor=blue,CJKbookmarks=true]{hyperref}
\usepackage{amsmath,amsfonts,amssymb}
\usepackage{graphicx}
\usepackage{color}
\usepackage[dvipsnames,svgnames,x11names]{xcolor}
\usepackage{slashed} 
\usepackage{url}
\usepackage{subfigure}
\usepackage{multirow} 
\usepackage{setspace} 
\graphicspath{{figs/}}  
\makeatother
\allowdisplaybreaks 

\begin{document}

\setstretch{1.2} 

\title{Improving cosmological parameter estimation with the future 21 cm observation from SKA }
\author{Jing-Fei Zhang$^1$}
\author{Li-Yang Gao$^1$}
\author{Dong-Ze He$^1$}
\author{Xin Zhang$^{1,2,3}$}\email{zhangxin@mail.neu.edu.cn}
\date{\today}
\affiliation{$^1$Department of Physics, College of Sciences, Northeastern University, Shenyang 110819, China}
\affiliation{$^2$Ministry of Education's Key Laboratory of Data Analytics and Optimization for Smart Industry, Northeastern University, Shenyang 110819, China}
\affiliation{$^3$Center for High Energy Physics, Peking University, Beijing 100080, China}

\begin{abstract}
Future observations of 21 cm emission from neutral hydrogen survey will become a promising approach to probe the large scale structure of the Universe.
In this paper, we investigate the impacts of Square Kilometre Array (SKA) 21 cm observation on the estimation of cosmological parameters.
We use the simulated data of the baryonic acoustic oscillation (BAO) measurements based on the future SKA experiment with the intensity mapping (IM) technique to do the analysis. For the current observations, we use the latest cosmic microwave background (CMB) observation from {\it Planck}, the optical BAO measurements, and the Type Ia supernovae (SN) observation (Pantheon compilation).
We find that the SKA mock data could break the degeneracy between the matter density and the Hubble constant, further improving the cosmological constraints to a great extent.
We also find that the constraint on the equation of state parameters of dark energy could be significantly improved by including the SKA mock data into the cosmological global fit.

\end{abstract}

\maketitle

\section{Introduction}
During the past decades, the standard cosmology has been established thanks to a series of accurate cosmological observations, such as the cosmic microwave background (CMB) anisotropies measured by {\it Planck} satellite \cite{Aghanim:2018eyx,Aghanim:2015xee,Ade:2015xua,Ade:2013zuv}, the distance-redshift relation measurements from Type Ia supernovae (SN) \cite{Riess:1998cb,Perlmutter:1998np}, and the large scale structure survey from Sloan Digital Sky Survey (SDSS) and six-degree-field Galaxy Survey (6dFGS) \cite{Reid:2009xm,Reid:2012sw,Ross:2014qpa,Beutler:2011hx} etc. The energy budget of current Universe consists of {5\%} baryonic matter, {27\%} dark matter and {68\%} dark energy. Unveiling the mystery of dark sectors has become the most fundamental mission of modern cosmology and physics.

For dark energy (DE), if it is responsible for the cosmic accelerated expansion, the most obvious theoretical candidate is the cosmological constant (vacuum energy) $\Lambda$ which has the equation of state (EoS) $w=-1$. Thus far, the cosmological constant scenario (also {known} as $\Lambda$CDM model) serves as the prototype of {the standard model of} cosmology, as it is in excellent agreement with current cosmological observations with the {fewest} free parameters. However, as is well known, there are two theoretical problems resulting from the $\Lambda$CDM model, namely the ``fine-tuning" problem and the ``cosmic coincidence" problem. Despite that great efforts have been devoted to resolving these two difficulties, all the attempts turn out to be unsuccessful. Additionally, the $\Lambda$CDM model has also encountered serious challenges in the aspect of conflicts between various experimental observations such as the Hubble constant $H_{0}$ tension and so forth \cite{Ade:2015xua}. Given this circumstance, it is hard to believe that the cosmological constant model with only six primary parameters is the eventual scenario of our universe. Undoubtedly, any deviation from $\Lambda$CDM model, if confirmed, would be an outstanding breakthrough for cosmology.

In recent years, we have come into an era of precision cosmology. The CMB measurements from {\it Planck} mission have newly constrained the parameters within the standard $\Lambda$CDM model with unprecedented accuracy $\lesssim 1\%$ \cite{Aghanim:2018eyx}. However, for the study of DE which dominates the expansion of late time universe, one also requires the observations at much lower redshifts ($z\lesssim 1$) than CMB ($z\approx 1100$), which can be achieved by measuring the cosmic large scale structures (LSS). The baryon acoustic oscillation (BAO) is just an essential cosmological probe {extracted from} LSS, which is imprinted by the cosmic acoustic waves in the early universe with a well-known scale of the matter distribution at around 150 Mpc. Using BAO signatures in the matter power spectrum as a {\it standard ruler}, high precision measurements on the cosmological parameters such as the EoS parameter of DE would be obtained.

Conventionally, the measurement of BAO is usually achieved by the optical galaxy surveys that detect individual galaxies with high resolution. Notwithstanding, other than those optical surveys we also need to develop new types of cosmological probe in the forthcoming future. To measure the BAO, there is an alternative impressive approach at the radio wavelength through the 21 cm intensity mapping (IM) technique. The 21 cm emission line comes from the spin-flip transition of electrons in neutral hydrogen, and serves as a great tracer of matter density fluctuations in the universe.
Recently, a number of intensity mapping projects have been proposed, for example the BAO from Integrated Neutral Gas Observations (BINGO) project \cite{Battye:2012tg}, the Canadian Hydrogen Intensity Mapping Experiment (CHIME) telescope \cite{Bandura:2014gwa}, and the Tianlai (``heavenly sound" in Chinese) project \cite{Chen:2012xu}.

The IM facility considered in our analysis is the upcoming Square Kilometre Array (SKA), which will be delivered in two phases, with SKA Phase 1 (denoted as SKA1) currently being under construction and the configuration of SKA Phase 2 (denoted as SKA2) being designed \cite{Bacon:2018dui}. SKA1 is comprised of two telescopes, i.e., SKA1-MID and SKA1-LOW. SKA1-MID is a mid-frequency dish array located in South Africa, observing the radio frequency between 0.35 GHz $-$ 1.75 GHz, and SKA1-LOW is located in the western of Australia, observing between 0.05 GHz $-$ 0.35 GHz \cite{Bacon:2018dui}. Moreover, SKA1-MID will operate in two frequency bands, with Band1 observing at 350 MHz $-$ 1050 MHz and Band2 observing at 1050 MHz $-$ 1750 MHz.  Furthermore, SKA2 will perform an immense galaxy redshift survey over three quarters of the sky, with an impressive sensitivity surpassing almost all the other planned BAO measurements at the redshift range $0.4 \lesssim z \lesssim 1.3$.  We primarily concentrated on SKA1-MID and SKA2 in this work, since SKA1-MID contains the main frequency range for IM and they are both able to observe LSS at low redshifts $0\lesssim z \lesssim 3$ where DE dominates the evolution of cosmos.

On the other hand, as is well known, astronomical observation is the key to determine the nature of DE. With the advancement in astronomical observation technology over the past decades, neutral hydrogen 21 cm radio cosmology has become one of the most essential breakthroughs we are dedicated to making. So, as an ambitious neutral hydrogen survey project, SKA will play an important role in the measurement of cosmological parameters and in the exploration of the nature of DE. For the investigations of the constraints on the EoS of DE using the mock data from SKA, see Refs. \cite{Bacon:2018dui,Bull:2015nra,Raccanelli:2015qqa,Zhao:2015wqa,Li:2019loh,Zhang:2019ipd}. In Ref. \cite{Bull:2015nra}, the authors have made a concrete simulation on the measurements of BAO in light of the 21 cm neutral hydrogen surveys from SKA1 and SKA2, and provided a relatively conservative estimation on the cosmological constraints. Hence, we can directly use these simulated data to do the analysis. In Ref.\;\cite{Zhang:2019ipd}, the authors utilized these mock data to investigate the prospects of weighing the mass of neutrinos in the $\Lambda$CDM universe with future SKA observations. In this work, we will use these SKA simulated data to perform constraints on the parameters in three cosmological models, i.e., the $\Lambda$CDM model, the $w$CDM model, and the Chevalliear-Polarski-Linder (CPL) model \cite{Chevallier:2000qy,Linder:2002et}, respectively.
The aim of this work is to assess the potential of future SKA project for improving the constraints on the cosmological parameters when including its simulated data in the cosmological global fit.


\section{Method and data}

We first {use} the current observational data to constrain different cosmological models. We choose the current mainstream cosmological probes, namely CMB, BAO, and SN. For the CMB data, we use the distance priors data from {\it Planck} 2018 \cite{Aghanim:2018eyx,Chen:2018dbv}. For the BAO data, we use the measurements from 6dFGS ($z_{\rm eff}=0.106$) \cite{Beutler:2011hx}, SDSS-MGS ($z_{\rm eff}=0.15$) \cite{Ross:2014qpa}, and BOSS DR12 ($z_{\rm eff}=0.38$, 0.51, and 0.61) \cite{Alam:2016hwk}. For the SN data, we use the latest Pantheon compilation, which is comprised of 1048 data points from the Pantheon compilation \cite{Scolnic:2017caz}. For convenience, the data combination ``CMB+BAO+SN" is abbreviated as ``CBS" in the following.
In order to constrain the cosmological parameters, we employ the MCMC package {\tt CosmoMC} \cite{Lewis:2002ah} to infer their posterior probability distributions, and further to derive their best-fit values and corresponding errors.

We use the simulated data of the BAO measurements from the neutral hydrogen sky survey based on SKA1 and SKA2 in Ref. \cite{Bull:2015nra}. The relative errors of cosmic expansion rate $\sigma_{H}/H$ and the angular diameter distance $\sigma_{D_A}/D_A$ can be directly extracted from Fig. 3 in Ref. \cite{Bull:2015nra}.
As for the SKA1 data, we separately use the simulated data from SKA1-MID Band1 containing 11 data points of $\sigma_{H}/H$ as well as 6 data points of $\sigma_{D_A}/D_A$, and use the simulated data from SKA1-MID Band2 with 7 data points of $\sigma_{H}/H$ and 8 data points of $\sigma_{D_A}/D_A$. With respect to the SKA2, we use the simulated data including 17 data points of $\sigma_{H}/H$ and 17 data points of $\sigma_{D_A}/D_A$. Then, the likelihood function can be naturally established using these data points. Besides the data of SKA, we also use the data simulated based on the future optical experiment Euclid, which contains 14 data points of $\sigma_{H}/H$ and 14 data points of $\sigma_{D_A}/D_A$. Euclid is a 1.2 m near-infrared space telescope that is under development by the European Space Agency (ESA) and is selected for launch in 2020.  Since the operations of Euclid and SKA1 will be carried out during roughly the same period, we also take this optical experiment into consideration to make a comparison. We consider five various data combinations in this work: CBS, CBS+Euclid, CBS+SKA1, CBS+SKA2, CBS+Euclid+SKA2.

We employ three typical and simple dark energy models to perform the analysis. Besides the $\Lambda$CDM model mentioned above, we also consider the $w$CDM model and the CPL model in this work. The $w$CDM model is the simplest extension to the $\Lambda$CDM model, in which the EoS parameter $w$ is not equal to $-1$, but assumed to be a constant instead. The CPL model is a further extension to the $\Lambda$CDM model. In this parameterization DE scenario, the EoS is assumed to be of the form $w(a)=w_{0}+w_{a}(1-a)$ to describe the cosmological evolution of DE, with the scale factor $a\equiv 1/(1+z)$ and $w_{0}$ as well as $w_{a}$ being the free parameters to be constrained by experimental observations.

In a spatially flat universe (the assumption of flatness is motivated by the inflation scenario and, actually, is strongly favored by current observations), the Hubble expansion rate shall be given by the Friedmann equation. 
In the $\Lambda$CDM model, the Friedmann equation reads
\begin{equation*}
E^{2}(z)=\Omega_{m}(1+z)^{3}+\Omega_{r}(1+z)^{4}+(1-\Omega_{m}-\Omega_{r}),
\end{equation*}
with $E(z)\equiv H(z)/H_{0}$. Here, $\Omega_m$ and $\Omega_r$ denote the fractional densities of matter and radiation, respectively.
In the $w$CDM model, the Friedmann equation is of the form
\begin{multline*}
E^{2}(z)=\Omega_{m}(1+z)^{3}+\Omega_{r}(1+z)^{4}\\
+(1-\Omega_{m}-\Omega_{r})(1+z)^{3(1+w)}.
\end{multline*}
In the CPL model, the Friedmann equation is
\begin{multline*}
E^{2}(z)=\Omega_{m}(1+z)^{3}+\Omega_{r}(1+z)^{4} \\
+(1-\Omega_{m}-\Omega_{r})(1+z)^{3(1+w_{0}+w_{a})}{\rm exp}\,\left(-\frac{3w_{a}z}{1+z}\right).
\end{multline*}
The angular diameter distance $D_{A}(z)$ can be calculated through the formula $D_{A}(z)=(1+z)^{-1}\int^{z}_{0}dz^{\prime}/H(z^{\prime})$ from a specific cosmological model.

\section{Results and discussion}

\begin{figure*}[!htp]
\includegraphics[width=0.6\textwidth]{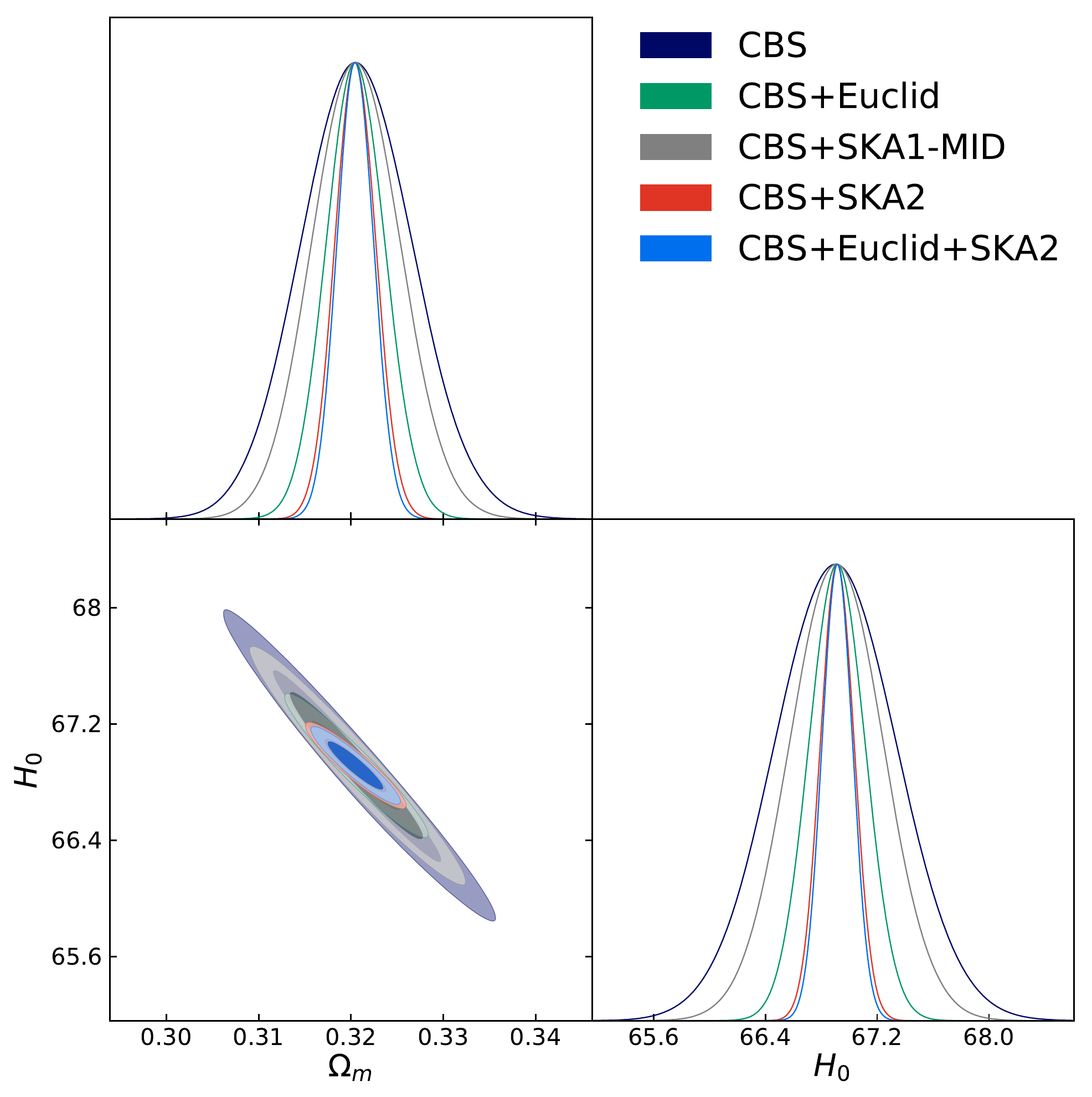}
\caption{Observational constraints ($68.3\%$ and $95.4\%$ confidence level) on the ${\rm \Lambda CDM}$ model by using the ${\rm CBS}$, ${\rm CBS+Euclid}$, ${\rm CBS+SKA1}$, ${\rm CBS+SKA2}$ and ${\rm CBS+Euclid+SKA2}$ data combinations. Here, ${\rm CBS}$ stands for ${\rm CMB+BAO+SN}$ and $H_0$ is in units of ${\rm km\ s^{-1}\ Mpc^{-1}}$.}
\label{fig1}
\end{figure*}

\begin{figure*}
\includegraphics[width=0.3\textwidth]{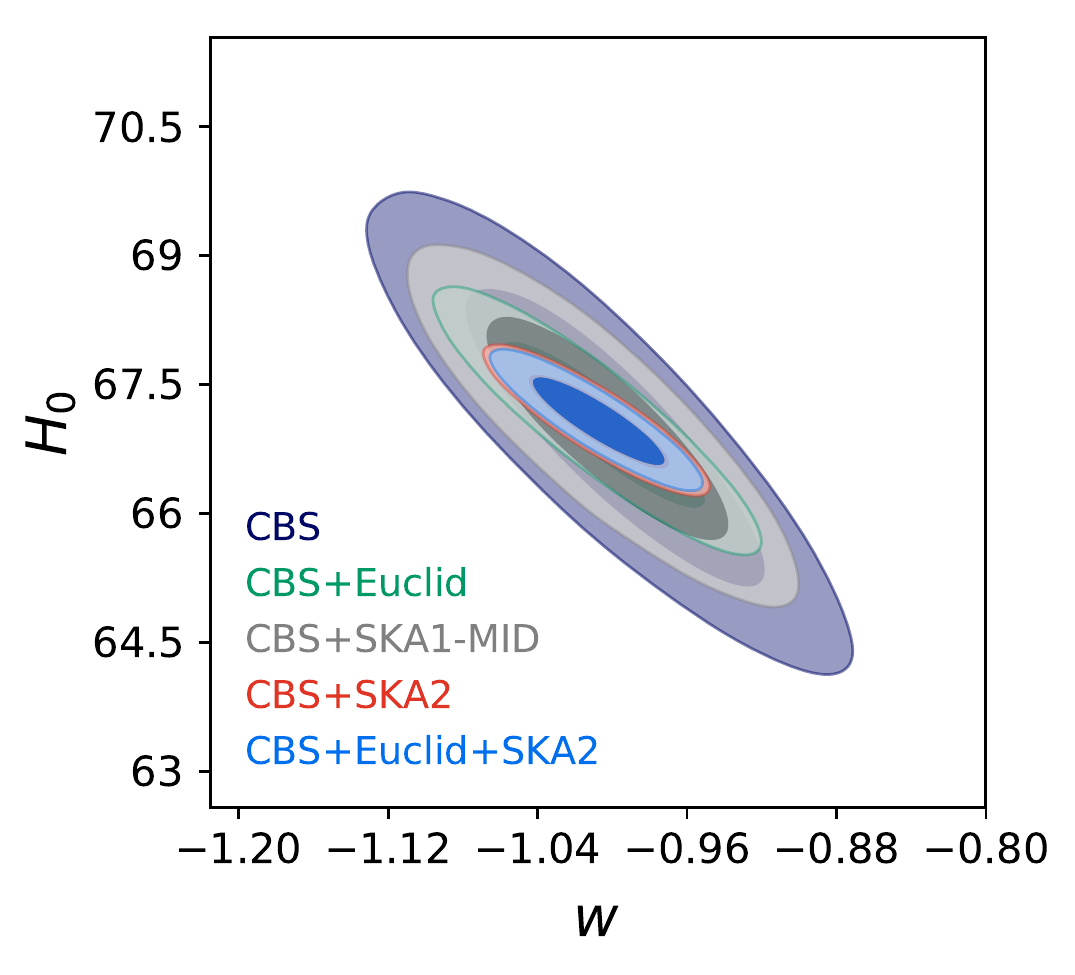}
\includegraphics[width=0.3\textwidth]{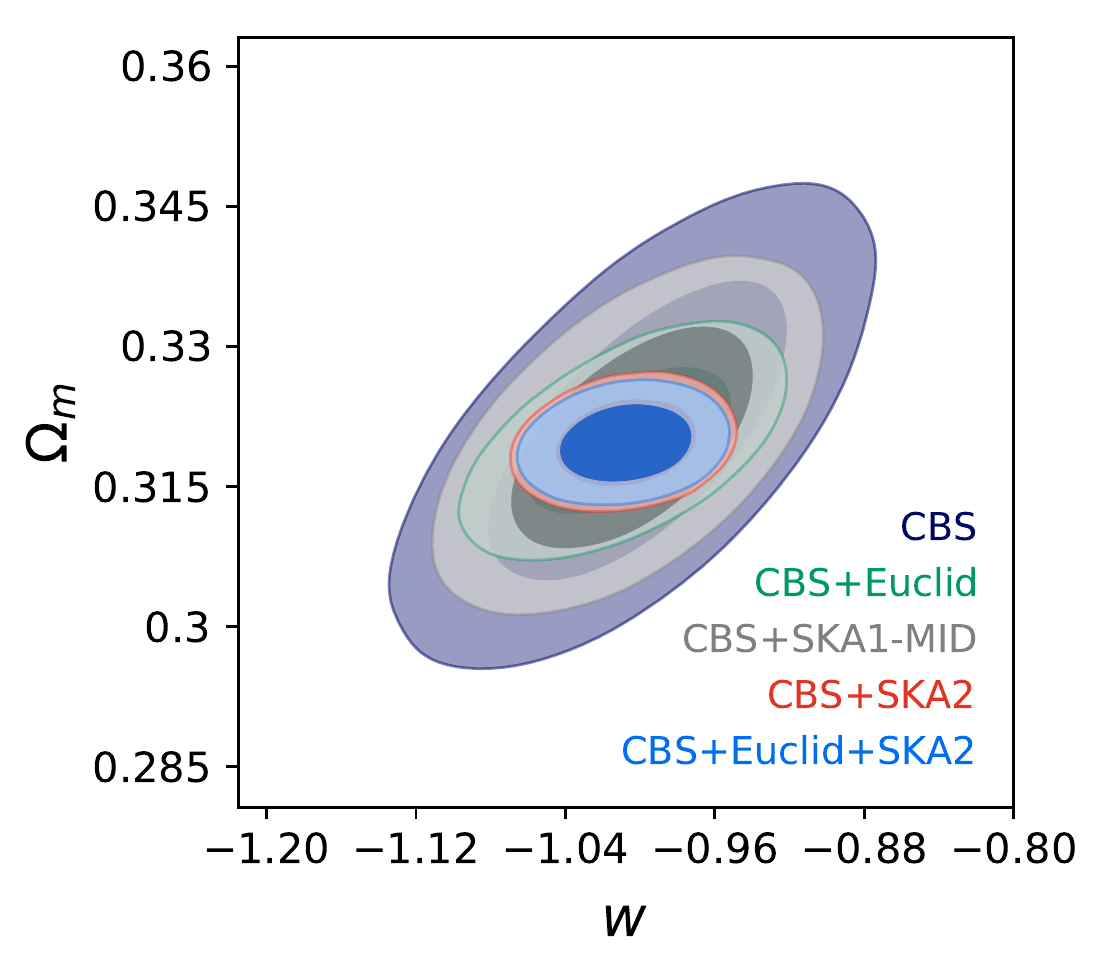}
\includegraphics[width=0.3\textwidth]{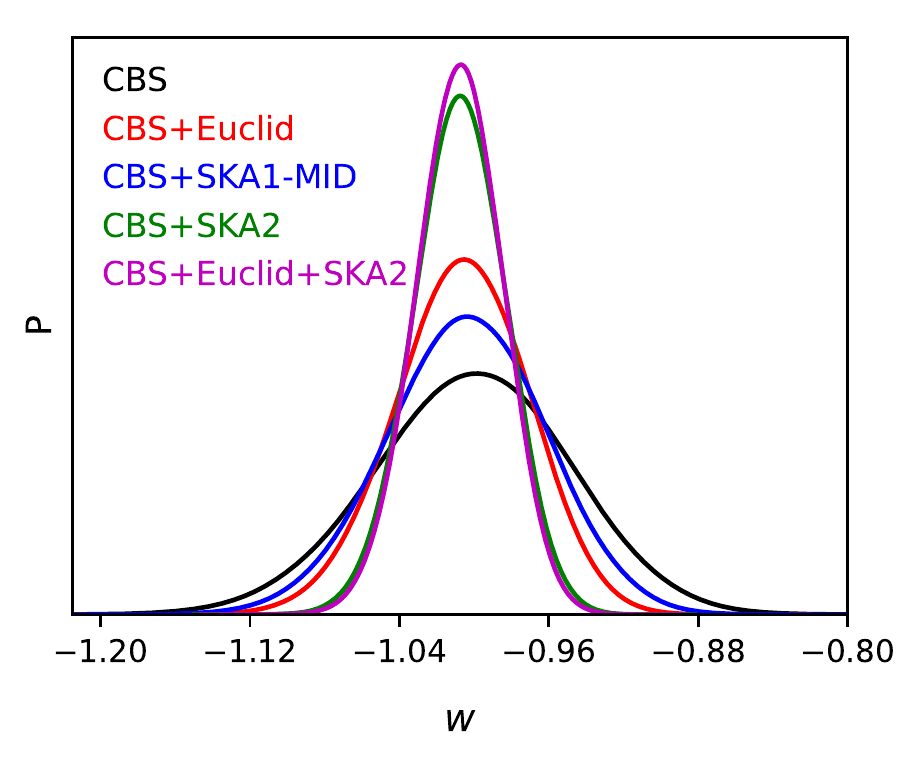}
\caption{Observational constraints ($68.3\%$ and $95.4\%$ confidence level) on the ${ w\rm CDM}$ model by using the ${\rm CBS}$, ${\rm CBS+Euclid}$, ${\rm CBS+SKA1}$, ${\rm CBS+SKA2}$ and ${\rm CBS+Euclid+SKA2}$ data combinations. Here, ${\rm CBS}$ stands for ${\rm CMB+BAO+SN}$ and $H_0$ is in units of ${\rm km\ s^{-1}\ Mpc^{-1}}$.}
\label{fig2}
\end{figure*}

\begin{figure*}
\includegraphics[width=0.5\textwidth]{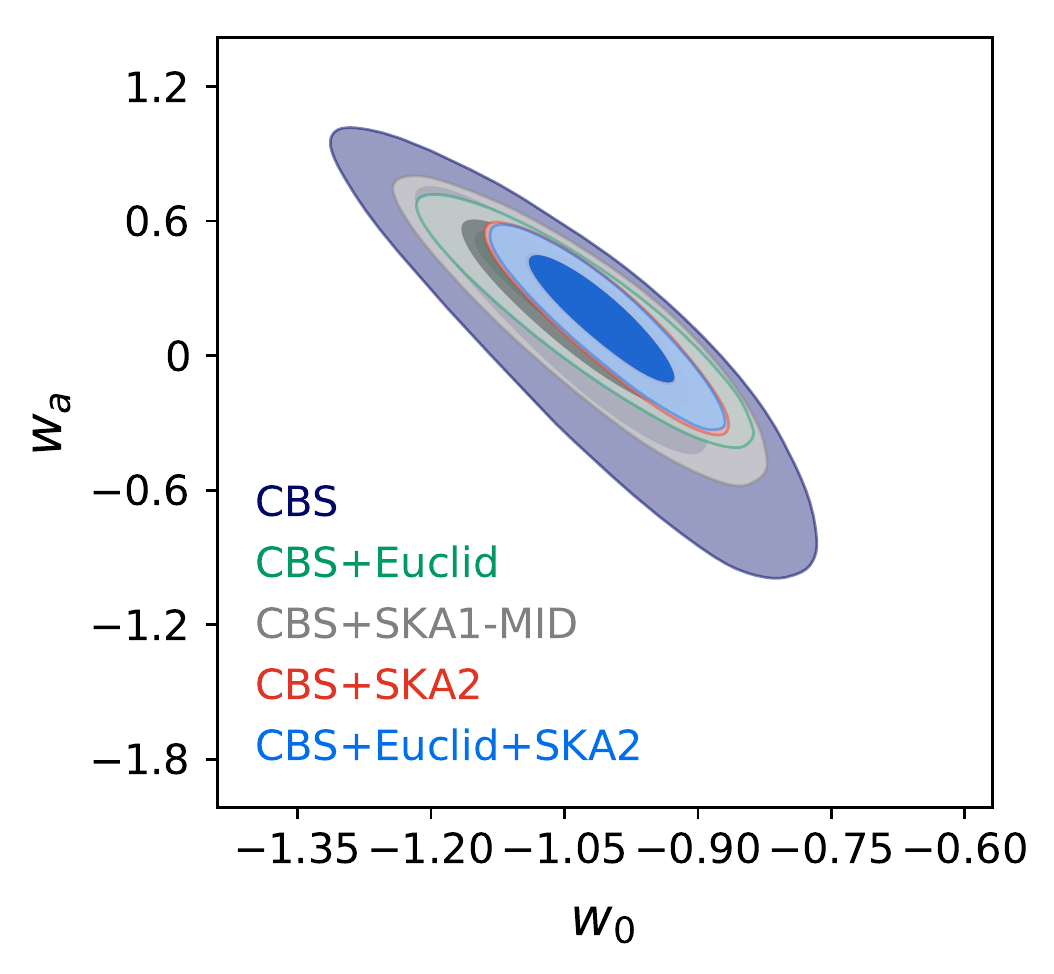}
\caption{Observational constraints ($68.3\%$ and $95.4\%$ confidence level) on the ${\rm CPL}$ model by using the ${\rm CBS}$, ${\rm CBS+Euclid}$, ${\rm CBS+SKA1}$, ${\rm CBS+SKA2}$ and ${\rm CBS+Euclid+SKA2}$ data combinations. Here, ${\rm CBS}$ stands for ${\rm CMB+BAO+SN}$.}
\label{fig3}
\end{figure*}

In this section, we shall report the constraint results and make some relevant discussions. We constrain the ${\rm \Lambda CDM}$, ${ w\rm CDM}$, and ${\rm CPL}$ models by using the data combinations of CBS, CBS+Euclid, CBS+SKA1, CBS+SKA2, and CBS+Euclid+SKA2 to complete our analysis. Our main constraint results for cosmological parameters are shown in Figs.~\ref{fig1}--\ref{fig3} and summarized in Tables~\ref{tab1}--\ref{tab3}. In Figs.~\ref{fig1}--\ref{fig3}, we display the two-dimensional posterior distribution contours for various model parameters constrained at 68\% and 95\% confidence level. In Tables~\ref{tab1}--\ref{tab3}, we exhibit the best fitting values with 1$\sigma$ errors quoted, the constraint errors and the constraint accuracies of concerned parameters (i.e., $\Omega_{m}$, $H_{0}$, $w$, $w_{0}$, and $w_{a}$) in different cosmological models. In the tables, $\sigma(\xi)$ is the 1$\sigma$ error of the parameter $\xi$, and the constraint precision $\varepsilon (\xi)$ for the parameter $\xi$ is defined as $\varepsilon (\xi)=\sigma(\xi)/\xi_{\rm bf}$, where $\xi_{\rm bf}$ represents the best-fit value.


From Figs.~\ref{fig1}--\ref{fig3}, we can clearly find that future Euclid, SKA1, and SKA2 observations can significantly improve the constraints on almost all the parameters to some different extent, as shown by the green, gray, and red contours. Compared with SKA1 and Euclid data, in particular, the SKA2 data possess a much more powerful constraint capability; for more details, see also Tables~\ref{tab1}--\ref{tab3}. In Fig.\;\ref{fig1}, we find that the SKA mock data could break the degeneracy between the matter density and the Hubble constant, and further improve the cosmological constraints to a great extent, which is consistent with the conclusion in Ref. \cite{Zhang:2019ipd}. Meanwhile, in Fig.\;\ref{fig2}, we also find that the parameter degeneracy orientations of CBS+SKA2 and CBS+Euclid evidently differ from that of CBS only data combination as shown in the $\Omega_{m}$--\,$w$ plane. In other words, the Euclid and SKA2 mock data can help to break the parameter degeneracies, in particular between the parameters $\Omega_{m}$ and $w$ in the $w$CDM model. In Fig.\;\ref{fig3}, we also show the constraint results on the model parameters $w_0$ and $w_a$, and we find that the constraining capability of SKA is still powerful, especially for the case of SKA2. Apart from that, in all these figures we can apparently see that, the data of Euclid behave much better than SKA1 but worse than SKA2 in the parameter constraints, which indicates that future optical experiment Euclid will become more powerful in parameter constraints than the contemporaneous experiment, the first phase of SKA (SKA1). But with the development of SKA,  in its second phase (SKA2) this project will still become the most competitive experiment. Thus, in the following, we will focus the discussion on the SKA2 mock data.

When adding the SKA2 mock data to the current optical observations, the CBS datasets, the improvement of the constraint on the parameter $\Omega_{m}$ is from 2.65\% to 1.00\% in the ${\rm \Lambda CDM}$ model, from 3.13\% to 0.94\% in the  ${ w\rm CDM}$ model, from 3.32\% to 2.19\% in the ${\rm CPL}$ model. For the parameter $H_{0}$, the constraint precision is improved from 0.93\% to 0.25\% in the ${\rm \Lambda CDM}$ model, from 1.64\% to 0.52\% in the ${ w\rm CDM}$ model, and from 1.59\% to 0.77\% in the ${\rm CPL}$ model. With respect to the parameters featuring the property of DE, the improvement is from 5.18\% to 2.33\% for the parameter $w$ in the ${ w\rm CDM}$ model, and from 9.66\% to 5.16\% for the parameter $w_{0}$ in the CPL model. It should be mentioned that, since the central value of the parameter $w_a$ in the CPL model is around zero, the relative error of this parameter will be immensely influenced by the statistic fluctuations. Hence, the absolute error for this parameter is more reliable for quantifying the improvement. For this parameter, the absolute constraint error is improved from 0.3646 to 0.1836.

When adding the SKA2 mock data to the data combination of CBS+Euclid, the improvement of the constraint on the parameter $\Omega_{m}$ is from 1.55\% to 0.84\% in the ${\rm \Lambda CDM}$ model, from 0.90\% to 0.48\% in the  ${ w\rm CDM}$ model, from 2.83\% to 2.16\% in the ${\rm CPL}$ model. For the parameter $H_{0}$, the constraint precision is improved from 0.43\% to 0.24\% in the ${\rm \Lambda CDM}$ model, from 0.90\% to 0.48\% in the ${ w\rm CDM}$ model, and from 1.23\% to 0.72\% in the ${\rm CPL}$ model. With regard to the parameters featuring the property of DE, the improvement is from 3.40\% to 2.16\% for the parameter $w$ in the ${ w\rm CDM}$ model, and from 7.17\% to 4.88\% for the parameter $w_{0}$ in the CPL model. For the parameter $w_a$, the absolute error is improved from 0.2213 to 0.1761, and the improvement is significant as well.
Therefore, we can conclude that the BAO measurements from SKA with 21 cm IM technique will be able to significantly improve the cosmological parameter constraints in the future.

\begin{table*}
\renewcommand{\arraystretch}{1.5}
\caption{\label{tab1}The best-fit values of parameters, corresponding errors and constraint precisions based on the ${\rm \Lambda CDM}$ model by using CBS, CBS+Euclid, CBS+SKA1, CBS+SKA2 and CBS+Euclid+SKA2 data combinations. Here, CBS stands for CMB+BAO+SN and $H_0$ is in units of ${\rm km\ s^{-1}\ Mpc^{-1}}$.}
\begin{tabular}{|c|c c c c c|}
\hline
&CBS&CBS+Euclid&CBS+SKA1&CBS+SKA2&CBS+Euclid+SKA2\\
\hline
$H_0$&$66.92^{+0.62}_{-0.61}$&$66.91\pm0.29$&$66.91^{+0.48}_{-0.47}$&$66.92\pm0.17$&$66.92^{+0.15}_{-0.16}$\\
 $\Omega_{\rm m}$&$0.3205^{+0.0086}_{-0.0084}$&$0.3205\pm0.0045$&$0.3205^{+0.0068}_{-0.0067}$&$0.3204^{+0.0032}_{-0.0031}$&$0.3204\pm0.0028$\\

\hline
  $\sigma(H_0)$&$0.62$&$0.29$&$0.48$&$0.17$&$0.16$\\
 $\sigma(\Omega_{\rm m})$&$0.0085$&$0.0045$&$0.0068$&$0.0032$&$0.0028$\\
\hline
$\varepsilon(H_0)$&$0.0093$&$0.0043$&$0.0072$&$0.0025$&$0.0024$\\
 $\varepsilon(\Omega_{\rm m})$&$0.0265$&$0.0140$&$0.0212$&$0.0100$&$0.0087$\\
\hline
\end{tabular}
\end{table*}

\begin{table*}
\renewcommand{\arraystretch}{1.5}
\caption{\label{tab2}Same as  Table \ref{tab1}, but for the $w$CDM model (adding one more parameter, $w$).
}
\begin{tabular}{|c|c c c c c|}
\hline
&CBS&CBS+Euclid&CBS+SKA1&CBS+SKA2&CBS+Euclid+SKA2\\
\hline
$w$&$-1.0084^{+0.0599}_{-0.0431}$&$-1.0026^{+0.0316}_{-0.0365}$&$-1.0077^{+0.0450}_{-0.0366}$&$-1.0073^{+0.0238}_{-0.0233}$&$-1.0060^{+0.0204}_{-0.0229}$\\
  $H_0$&$67.09^{+1.01}_{-1.19}$&$67.01^{+0.63}_{-0.56}$&$67.05^{+0.76}_{-0.88}$&$67.09^{+0.33}_{-0.37}$&$67.06^{+0.34}_{-0.29}$\\
 $\Omega_{\rm m}$&$0.3196^{+0.0108}_{-0.0092}$&$0.3198^{+0.0050}_{-0.0049}$&$0.3202^{+0.0073}_{-0.0075}$&$0.3195^{+0.0032}_{-0.0027}$&$0.3197^{+0.0025}_{-0.0028}$\\

\hline
  $\sigma(w)$&$0.0522$&$0.0341$&$0.0410$&$0.0235$&$0.0217$\\
  $\sigma(H_0)$&$1.10$&$0.60$&$0.82$&$0.35$&$0.32$\\
 $\sigma(\Omega_{\rm m})$&$0.0100$&$0.0050$&$0.0074$&$0.0030$&$0.0027$\\
 \hline
$\varepsilon(w)$&$0.0518$&$0.0340$&$0.0407$&$0.0233$&$0.0216$\\
$\varepsilon(H_0)$&$0.0164$&$0.0090$&$0.0122$&$0.0052$&$0.0048$\\
 $\varepsilon(\Omega_{\rm m})$&$0.0313$&$0.0156$&$0.0231$&$0.0094$&$0.0084$\\
\hline
\end{tabular}
\end{table*}

\begin{table*}
\renewcommand{\arraystretch}{1.5}
\caption{\label{tab3}Same as Table \ref{tab1}, but for the ${\rm CPL}$ model (adding two more parameters, $w_{0}$ and $w_{a}$).
}
\begin{tabular}{|c|c c c c c|}
\hline
&CBS&CBS+Euclid&CBS+SKA1&CBS+SKA2&CBS+Euclid+SKA2\\
\hline
$w_0$&$-1.0644^{+0.1102}_{-0.0948}$&$-1.0332^{+0.0684}_{-0.0793}$&$-1.0447^{+0.0873}_{-0.0791}$&$-1.0179^{+0.0581}_{-0.0462}$&$-1.0103^{+0.0492}_{-0.0494}$\\
 $w_a$&$0.2548^{+0.3397}_{-0.3879}$&$0.2365^{+0.2169}_{-0.2258}$&$0.2303^{+0.2438}_{-0.2504}$&$0.2030^{+0.1631}_{-0.2020}$&$0.1718^{+0.1873}_{-0.1642}$\\
  $H_0$&$67.20^{+0.90}_{-1.22}$&$66.55^{+0.96}_{-0.68}$&$66.72\pm0.87$&$66.36^{+0.46}_{-0.55}$&$66.32\pm0.48$\\
 $\Omega_{\rm m}$&$0.3161^{+0.0123}_{-0.0084}$&$0.3218^{+0.0084}_{-0.0098}$&$0.3210^{+0.0089}_{-0.0091}$&$0.3240^{+0.0072}_{-0.0070}$&$0.3246^{+0.0068}_{-0.0071}$\\

\hline
  $\sigma(w_0)$&$0.1028$&$0.0741$&$0.0833$&$0.0525$&$0.0493$\\
  $\sigma(w_a)$&$0.3646$&$0.2213$&$0.2471$&$0.1836$&$0.1761$\\
  $\sigma(H_0)$&$1.07$&$0.83$&$0.87$&$0.51$&$0.48$\\
  $\sigma(\Omega_{\rm m})$&$0.0105$&$0.0091$&$0.0090$&$0.0071$&$0.0070$\\
\hline
$\varepsilon(w_0)$&$0.0966$&$0.0717$&$0.0797$&$0.0516$&$0.0488$\\
$\varepsilon(w_a)$&$1.4309$&$0.9357$&$1.0729$&$0.9044$&$1.0250$\\
$\varepsilon(H_0)$&$0.0159$&$0.0125$&$0.0130$&$0.0077$&$0.0072$\\
  $\varepsilon(\Omega_{\rm m})$&$0.0332$&$0.0283$&$0.0280$&$0.0219$&$0.0216$\\
\hline
\end{tabular}
\end{table*}

\section{Conclusion}
In this paper, we evaluated the constraining power of future SKA 21 cm observations not only on the expansion related cosmological parameters $\Omega_{m}$ and $H_{0}$, but also on the phenomenological dynamical dark energy parameters (such as $w_{0}$ and $w_{a}$). We use the simulated data of the BAO measurements from the neutral hydrogen survey based on SKA1 and SKA2 to do the analysis. Besides, we also consider the simulation of future optical sky survey project Euclid as a comparison. For the current observations, we use the latest cosmic microwave background (CMB) observation from {\it Planck} 2018, the optical BAO measurements, and the Type Ia supernovae (SN) observation (Pantheon compilation).  We consider three popular cosmological models in this work, i.e., the $\Lambda$CDM model, $w$CDM model, and CPL model. We utilize five various data combinations: CBS, CBS+Euclid, CBS+SKA1, CBS+SKA2, CBS+Euclid+SKA2, to perform the constraints and to further make a comparison.

We find that SKA2 mock data will behave best in the parameter constraints in all the DE models. Regardless of adding the SKA2 mock data to either the CBS datasets or the CBS+Euclid datasets, the constraint results could be significantly improved in all the considered DE models. For example, with the addition of SKA2 in CBS datasets, the constraints on $\Omega_{m}$ can be improved by 34\%--70\%, and the constraints on $H_{0}$ can be improved by 52\%--73\%. For the constraints on those EoS parameters, the improvements are also obviously, with the parameter $w$ in $w$CDM promoted by 55.0\%, $w_0$ and $w_a$ in CPL model promoted by 46.6\% and 49.6\% respectively. In addition, we also find that the degeneracy between several cosmological parameters, such as $\Omega_m$ and $w$ in $w$CDM model as well as $\Omega_m$ and $H_0$ in $\Lambda$CDM model, could be effectively broken by the combination of Euclid and SKA mock data in the cosmological fit.

All in all, we conclude that in the future the 21 cm observation of SKA would be helpful to improve the parameter constraints on DE and have a great potential to become one of the most competitive cosmological radio probes to explore the property of DE.
\begin{acknowledgments}
This  work  was  supported  by  the  National  Natural Science  Foundation  of  China  (Grants  Nos. 11975072,  11875102, 11835009, 11690021, and 11522540)  and  the  National Program for Support of Top-Notch Young Professionals.
\end{acknowledgments}


\end{document}